\documentclass[superscriptaddress,twocolumn]{revtex4-2}

\usepackage[utf8]{inputenc}
\usepackage[english]{babel}
\usepackage{xcolor}
\usepackage{tikz}
\usepackage[colorlinks,urlcolor=blue,citecolor=blue,linkcolor=blue]{hyperref}
\usepackage{verbatim}
\usepackage{amsmath}
\usepackage[compress]{cleveref}
\usepackage{amssymb}
\usepackage{graphicx}
\usepackage{epstopdf}
\usepackage{esint}
\usepackage{lipsum}
\usepackage{braket}
\usepackage{textcomp}
\usepackage{gensymb}
\usepackage{booktabs}
\usepackage{setspace}
\usepackage{hyperref}

\newcommand{\termSym}[3]{\ensuremath{^{#1}\mathrm{#2}_{#3}}}

\newcommand\equip[1]{ (#1)}

\newcommand{\qtyunit}[2]{#1\,\text{#2}}

\begin{document}

\title{A Robust Strontium Tweezer Apparatus for Quantum Computing}%

\newcommand{\tue}{Coherence and Quantum Technology group, Department of Applied Physics and Science Education, Eindhoven University of Technology, P.O. Box 513, 5600 MB Eindhoven, The Netherlands}
\newcommand{\qte}{Center for Quantum Materials and Technologies Eindhoven (QT/e), P.O. Box 513, 5600 MB Eindhoven, The Netherlands}
\newcommand{\uva}{Van der Waals-Zeeman Institute, Institute of Physics, University of Amsterdam,
Science Park 904, 1098 XH Amsterdam, The Netherlands}
\newcommand{\qusoft}{QuSoft, Science Park 123, 1098 XG Amsterdam, The Netherlands}

\author{Marijn Venderbosch}
\thanks{These authors contributed equally to this work.}
\affiliation{\tue}
\affiliation{\qte}

\author{Rik van Herk}
\thanks{These authors contributed equally to this work.}
\affiliation{\tue}
\affiliation{\qte}

\author{Zhichao Guo}
\thanks{These authors contributed equally to this work.}
\affiliation{\tue}
\affiliation{\qte}

\author{Jesús del Pozo Mellado}
\affiliation{\tue}
\affiliation{\qte}

\author{Max Festenstein}
\affiliation{\tue}
\affiliation{\qte}

\author{Deon Janse van Rensburg}
\affiliation{\tue}
\affiliation{\qte}

\author{Ivo Knottnerus}
\affiliation{\tue}
\affiliation{\qte}
\affiliation{\uva}
\affiliation{\qusoft}

\author{Yu Chih Tseng}
\affiliation{\uva}
\affiliation{\qusoft}

\author{Alexander Urech}
\affiliation{\uva}
\affiliation{\qusoft}

\author{Robert Spreeuw}
\affiliation{\uva}
\affiliation{\qusoft}

\author{Florian Schreck}
\affiliation{\uva}
\affiliation{\qusoft}

\author{Rianne Lous}
\email[Author to whom correspondence should be addressed: ]{r.s.lous@tue.nl}
\affiliation{\tue}
\affiliation{\qte}

\author{Edgar Vredenbregt}
\affiliation{\tue}
\affiliation{\qte}

\author{Servaas Kokkelmans}
\affiliation{\tue}
\affiliation{\qte}

\date{January 23, 2026}

\begin{abstract}
    Neutral atoms for quantum computing applications show promise in terms of scalability and connectivity. 
We demonstrate the realization of a versatile apparatus capable of stochastically loading a $5\times5$ array of optical tweezers with single \textsuperscript{88}{Sr} atoms featuring flexible magnetic field control and excellent optical access.
A custom-designed oven, spin-flip Zeeman slower, and deflection stage produce a controlled flux of Sr directed to the science chamber.
In the science chamber, featuring a vacuum pressure of $3\times10^{-11}$ mbar, the Sr is cooled using two laser cooling stages, resulting in $\sim 3\times10^5$ atoms at a temperature of $5(1)$\,$\mu$K.
The optical tweezers feature a $1/e^2$ waist of $0.81(2)\,\mu\text{m}$, and loaded atoms can be imaged with a fidelity of $\sim0.997$ and a survival probability of $0.99^{+0.01}_{-0.02}$.
The atomic array presented here forms the core of a full-stack quantum computing processor targeted for quantum chemistry computational problems. 
\end{abstract}

\maketitle


\section{Introduction} 
    
    Ultracold atoms trapped in arrays of optical tweezers are a promising platform for quantum computing and quantum simulation \cite{chinnarasu_variational_2025, henriet_quantum_2020, dalyac_graph_2024}, due to their excellent connectivity \cite{bluvstein_logical_2024} and scalability \cite{manetsch_tweezer_2025, chiu_continuous_2025}.
    Recently, strontium (Sr) has emerged as an attractive candidate for such tweezer-based platforms \cite{norcia_microscopic_2018,cooper_alkaline-earth_2018}, driven by its favorable electronic structure that provides access to narrow and ultra-narrow optical transitions utilized in precision metrology \cite{takamoto_optical_2005,aeppli_clock_2024}.
    In Sr, the ultra-narrow clock transition can be used for qubit encoding \cite{madjarov_atomic-array_2019,norcia_microscopic_2018}, while the Rydberg-S states are accessible via a single-photon transition, offering high Rabi frequencies for high-fidelity entanglement gates \cite{madjarov_high-fidelity_2020, schine_long-lived_2022}. 

    Unlike alkali atoms, Sr exhibits low vapor pressure, which requires a high temperature oven \cite{schioppo_compact_2012} and multiple stages of laser cooling to be implemented in the experimental apparatus.
    Additionally, multiple laser systems are needed to address broad, narrow and ultra-narrow atomic transitions. 
    Over the years, several Sr setups have been built, demonstrating e.g. continuous loading of a magneto-optical trap (MOT) \cite{bennetts_steady-state_2017}, single-atom preparation \cite{cooper_alkaline-earth_2018,giardini_single_2025,wen_apparatus_2024}, motional ground state cooling \cite{holzl_motional_2023}, qubit manipulation \cite{norcia_microscopic_2018,ammenwerth_realization_2025} and entanglement  \cite{madjarov_high-fidelity_2020}.
    
    We have designed and realized a compact and versatile Sr apparatus capable of stochastically loading a $5\times5$ array of optical tweezers with single Sr atoms. 
    For robust operation, we opt for a deflection stage instead of the typically used 2-D MOT.
    The deflection stage based on transverse optical molasses directs a controlled flux of Sr atoms into the science chamber while removing the direct line-of sight between oven and science chamber.
    This allows us to maintain low vacuum pressure, which benefits long qubit coherence times, while still maintaining a sufficient loading rate for optical tweezer experiments.
    The deflection stage has a single retro-reflected laser beam that is forgiving with respect to beam alignment.
    For modularity, compactness, and stability purposes, the laser systems, referenced to a frequency comb, and the optics needed for intensity and frequency control are rack mounted, and the apparatus can run for multiple days without intervention.

    In this work, we demonstrate the successful operation of the apparatus for preparing single Sr atoms in tweezers. 
    Our results showcase the readiness of the setup to serve as a basis for single-qubit gate operations and Rydberg-mediated entanglement.
    In the near-term this apparatus will provide a neutral atom-based backend for the multi-hardware openly accessible platform Quantum Inspire \cite{noauthor_qutech_2025,noauthor_ry_2025}.
    We are implementing a full-stack approach and currently operate the Rydberg emulator RySP \cite{noauthor_tues_2025} as digital-twin of our apparatus.
    The paper is organized as follows.
    The laser systems involved are discussed in \cref{sec:lasers}, followed by the description of the apparatus in \cref{sec:exp_setup}.
    The successful demonstration of the apparatus, describing MOT operation, single atom preparation and imaging can be found in \cref{sec:single_atoms}.

\section{Laser systems}\label{sec:lasers} 

\begin{figure}
    \centering
    \includegraphics[width=\linewidth]{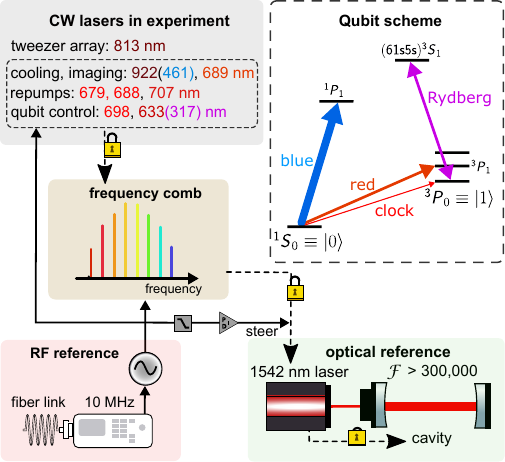}
    \caption{
        Laser system stabilization for the Sr-88 clock qubit.
        Seven continuous wave (CW) lasers are locked to a commercial frequency comb. 
        Light at 461 and \qtyunit{317}{nm} is generated by frequency doubling of the 922 and \qtyunit{633}{nm} fundamentals respectively, which are locked to the comb (doubled wavelengths in parentheses).
        The frequency comb is short-term stabilized to a high finesse cavity as optical reference and long-term steered using software-implemented feedback to an RF reference obtained from a fiber link to the Dutch Metrology Institute.
        The inset shows the energy level structure of Sr-88 and the atomic transitions addressed by the CW lasers. 
        Not shown: triplet-P transitions used for repumping.
    }
    \label{fig:lasers_overview}
\end{figure}

    The confinement, control, and detection of our strontium atoms depend on 8 continuous wave (CW) lasers with wavelengths ranging from $317\,\text{nm}$ to $813\,\text{nm}$, as shown in \cref{fig:lasers_overview}.
    The broad \termSym{1}{S}{0} $\to$ \termSym{1}{P}{1} transition ($461\,\text{nm}$) is used for initial cooling and imaging of Sr.
    A vertical-external-cavity surface-emitting (VECSEL) laser \equip{Vexlum, VALO SHG SF} and an injection-locked amplifier (ILA) system\equip{Moglabs} are used to produce in total $1.5\,\text{W}$ of $461\,\text{nm}$ light.
    The fundamental of the VECSEL laser at \qtyunit{922}{nm} is used for locking to the comb and is subsequently doubled to \qtyunit{461}{nm}.
    The $689\,\text{nm}$ light is produced by another ILA system\equip{Moglabs} to address the narrow \termSym{1}{S}{0} $\to$ \termSym{3}{P}{1} red transition.
    The optical tweezers are produced using a fiber laser system at $813\,\text{nm}$  \equip{Precilasers, FL-SF-813-8-CW}. 
    This wavelength is a magic wavelength for the \termSym{1}{S}{0} $\leftrightarrow$ \termSym{3}{P}{0} clock transition in Sr \cite{takamoto_optical_2005}, which we use as our qubit transition (\cref{fig:lasers_overview}).
    Optical pumping within the triplet $P$ states can be performed using three repump lasers\equip{Optoquest Co., Ltd.} coupling to the $(5s6s) \termSym{3}{S}{1}$ state.
    Laser light at $698\,\text{nm}$ \equip{Moglabs, ILA} is used for driving the clock transition. 
    Rydberg excitation is driven at \qtyunit{317}{nm} (Precilasers, FL-SF-316-1-CW), generated by second harmonic generation of the \qtyunit{633}{nm} fundamental locked to the comb. 
    The clock and Rydberg lasers are not discussed further in this work 

    All CW lasers except the $813\,\text{nm}$ laser are frequency stabilized using a commercial frequency comb.
    The frequency comb \equip{Menlo Systems, FC1500-250-ULN} is short-term stabilized by a beat lock to an ultra-stable $1542\,\text{nm}$ laser locked to a high-finesse cavity \equip{Menlo Systems, ORS mini}.
    The long-term drift of the cavity was measured to be $0.08919(7)\,\text{Hz/s}$.
    To remedy this, as well as thermal fluctuations, the comb is stabilized long-term using a 10 MHz RF signal \equip{Safran, WRS-LJ} obtained from a fiber link to the VSL Dutch Metrology Institute using the white rabbit protocol \cite{lipinski_white_2011}.
    For our clock and Rydberg lasers, which lock to comb index $m\sim 1.4\times 10^6$, this translates to a frequency stability of $\Delta f = m f_{\text{R}} \sigma(\tau) =\mathcal{O}(10)\,\text{Hz}$, using a modified Allan deviation $\sigma(\tau)\sim 4\times 10^{-13}$ after $\tau=10^4\,\text{s}$ of averaging the repetition rate $f_{\text{R}}=250\,\text{MHz}$.

    The laser system allows for continuous operation over several days or even weeks without intervention.
    The acousto-optic components and control electronics for laser detuning and intensity regulation are integrated into a 19-inch optical rack, providing a modular and space-efficient setup.
    The rack features a dust-proof sealing, an integrated ventilation system, and retractable $400 \times 700\,\text{mm}$ breadboards mounted on vibration-damping rubber posts.
    From the optical rack, all laser light is delivered to the vacuum chamber using optical fibers, with exception of the $317\,\text{nm}$ UV laser, which is positioned on the vacuum setup table.
    
\section{Experimental Setup}\label{sec:exp_setup} 

\begin{figure*}
    \centering
    \includegraphics[width=0.94\linewidth]{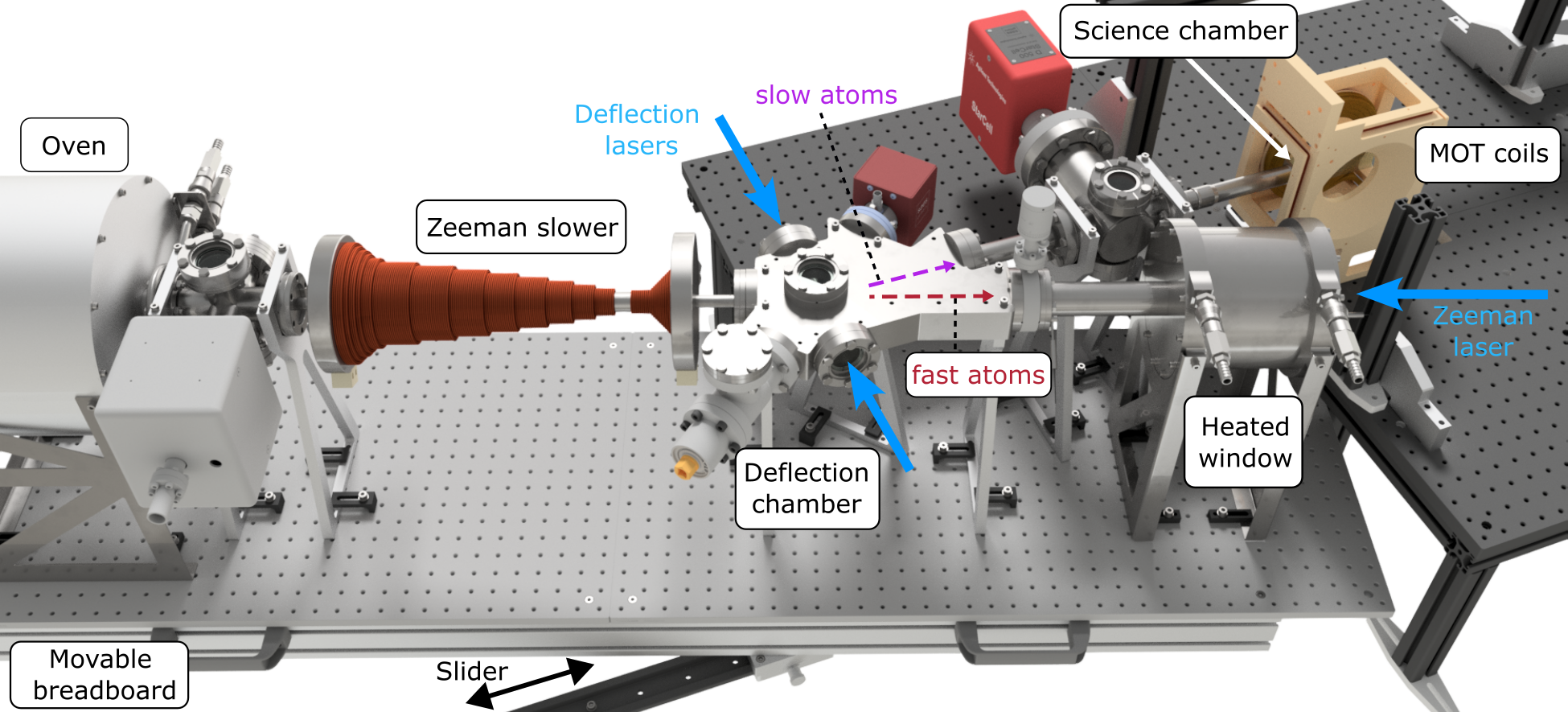}
    \caption{
        CAD image of the Sr tweezer apparatus.
        From the oven, the atoms are slowed in the spin-flip Zeeman slower using the Zeeman laser (light blue array).
        In the deflection chamber, the slowed atoms (purple dashed line), are deflected towards the science chamber using the deflection lasers (light blue arrows).
        The fast atoms (red dashed line) hit the heated window at $200\,\text{\degree}$.
        The science chamber consists of a glass cell mounted between MOT coils and microscope objectives (see \cref{fig:coils}).
    }
    \label{fig:overview_cad}
\end{figure*}

    A rendered image of the vacuum system is shown in \cref{fig:overview_cad}.
    It consists of three chambers, each pumped by a dedicated pump due to varying vacuum pressure requirements, that are connected by two differential pumping sections. 
    The vacuum system can be moved on a slider, so it can be removed when installing optics on the machine. 
    The optics for the machine is installed on the $19.05\,\text{mm}$ thick breadboards attached to $40\,\text{mm}$ profiling (black coloring). 
    The vacuum system is discussed below.

\subsection{Oven Chamber}

\begin{figure}
    \centering
    \includegraphics[width=\linewidth]{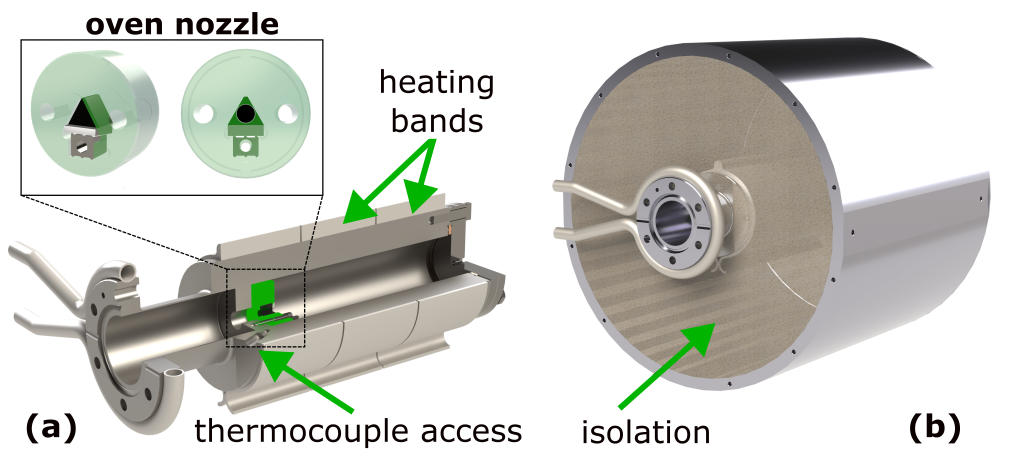}
    \caption{
        CAD images of the oven. 
        a) Cutout image of the inside of the oven, showing the heating bands, thermocouple access points and a zoomed image of the oven nozzle. 
        b) Oven enclosed in isolation material to avoid heat transferring to the rest of the vacuum system. 
    }
    \label{fig:oven_renders}
\end{figure}

    In the oven, Sr vapor is created by heating a Sr sample (Nova elements) to $420 \,\text{\degree C}$ to create a vapor pressure of $\sim 3\times10^{-4}\,\text{mbar}$ \cite{rumble_crc_2024}.
    The oven is based on a design described in \cite{bennetts_steady-state_2017} (supplemental material), which was inspired by Ref. \cite{senaratne_effusive_2015}.
    The design is shown in \cref{fig:oven_renders}, and focuses on generating a collimated atomic beam while providing thermal isolation from the rest of the vacuum chamber and the outside air. 

    The oven is a $69\,\text{mm}$ outer diameter stainless-steel cylinder (matching a CF40 flange) with a $36\,\text{mm}$ inner diameter.
    A nickel CF40 gasket connects it to the next vacuum section, chosen to avoid copper gasket reactivity and seal degradation at high temperatures.
    The chamber is pumped by an ion pump\equip{Agilent, StarCell 20}.

    The oven heat is provided by three heating bands\equip{Kurval, $300\,\text{W}$} that are tightly fastened around the oven.
    The temperature is monitored with a K-type thermocouple that is inserted into the wall of the oven as shown in \cref{fig:oven_renders}a.
    To prevent too much heat from escaping into the atmosphere around the rest of the experimental setup, the oven flange can be water cooled (\cref{fig:oven_renders}a) and the oven is surrounded by a larger hollow cylinder filled with glass wool for thermal insulation (\cref{fig:oven_renders}b.)

    To collimate the atomic beam, an oven nozzle is created by narrowing the inner diameter of the oven to $10\,\text{mm}$. 
    This hole is filled with 630 micro-tubes of $8\,\text{mm}$ in length and $0.165\,\text{mm}$ in inner diameter\equip{MicroGroup}, packed in a hexagonal lattice. 
    The expected atom flux of this design is $\approx 2\times 10^{13}\,\text{s\textsuperscript{-1}}$ with an angular full width at half maximum spread of $\approx 3.6$\,\text{\degree} 
\cite{jones_molecularbeam_1969,beijerinck_velocity_1975,steckelmacher_knudsen_1986}.

\subsection{Zeeman Slower}

    After the oven, the Sr atoms are decelerated by a Zeeman slower (ZS) before being captured in a blue MOT operating on the broad $461\,\text{nm}$ transition.
    The ZS consists of two water-cooled coils producing the magnetic field, wound around two coaxial tubes with the inner tube containing the vacuum and the space between the inner and outer tubes serving as a water-cooling channel.
    The inner channel also serves as a differential pumping stage. 
    To minimize the amount of cross-layer wires that cause unwanted extra field, the coils are wound in sets of two layers with equal windings wherever possible. 
    The first coil ($3.5\,\text{A}$) generates the field which generates the main part of the slowing, while the second coil ($6.5\,\text{A}$) creates a sharp reversal of the magnetic field.
    This spin-flip magnetic field configuration \cite{bennetts_steady-state_2017} was chosen for two reasons.
    A nonzero field at the end of the ZS allows atoms to drop out of resonance at the target velocity $v_t$.
    And a large initial field reduces the required detuning for a given capture velocity $v_0$. 

    The ZS design parameters were chosen on the basis of simulations and are summarized in \cref{tab:ZS_params}.
    The magnetic field maximum and capture velocity $v_0$ set the laser detuning of the counter-propagating $461\,\text{nm}$ laser beam (\cref{fig:overview_cad}) to $-420\,\text{MHz}$.
    The chosen capture velocity matches the mean thermal velocity at $420\,\text{°C}$. 
    Due to the more frequent decay of fast atoms via the intermediate $\termSym{1}{D}{2}$ state to the $^3$P$_{0,2}$ into metastable states, when slowed on the blue transition, a higher velocity would only increase the atom flux minimally. 
    The simulations predict an efficiency, defined by the fraction of atoms that exit the ZS below $v_t$, of  $\sim 1\,\%$.
    The rest of the atoms will have hit the walls of the vacuum chamber, left the cooling cycle, or left the oven with a velocity above the capture velocity.  

\begin{table}
    \caption{
        Design parameters of the Zeeman slower.
        The target B field length is the useful part of the slowing field. 
        The design parameter is the ratio of the deceleration compared to its maximum value. }
    \centering
    \begin{tabular}{lr}
        \hline\hline
        Tube diameter (mm)                  & 10    \\
        Tube length (mm)                    & 495   \\
        \hline
        Capture velocity $v_0$ (m/s)        & 422   \\
        Target velocity $v_t$ (m/s)         & 66    \\
        Laser detuning (MHz)                & -420  \\
        \hline
        B field start (G)                   & 345   \\
        B field end (G)                     & -197  \\
        Coil length (mm)                    & 350   \\
        Target B field length (mm)          & 300   \\
        \hline
        Design parameter                    & 0.3  \\
        \hline\hline
    \end{tabular}
    \label{tab:ZS_params}
\end{table}


\subsection{Deflection Chamber}

    Following the ZS, the Sr atoms enter the deflection chamber which deflects the slowed atoms to the science chamber and dumps the remaining atoms on a heated window.
    By deflecting only the atoms within the capture velocity of the MOT, the glass cell is less likely to be coated with Sr and a higher vacuum quality can be maintained.
    Additionally, due to the deflection, the line-of-sight between the oven and the science chamber is removed.
    This is anticipated to reduce the amount of black-body radiation, which has a negative effect on Rydberg state lifetime \cite{gallagher_interactions_1979}.
    The deflector also allows for quickly switching on and off of the atomic flux to the science chamber. 
    
    The deflection chamber is a commercial seven-port vacuum piece designed in-house. 
    The input port connects to the ZS and two output ports connect to the science chamber and heated window.
    Two more ports are used to connect a combined ion sputter and non-evaporative getter pump\equip{SAES Getters, NEXTorr Z100} and a valve to connect to an external pump.
    Finally, two view ports are mounted under a $70\,\text{\degree}$ angle with respect to the input port for transmission of the deflection laser beams.

    Using optical molasses, the slowed atoms are deflected by a \qtyunit{20}{\degree} angle with respect to the atom flux direction.
    For this, a retro-reflected deflection beam (\cref{fig:overview_cad}) of $461\,\text{nm}$ light with detuning of $-30\,\text{MHz}$, beam waist of $5.4\,\text{mm}$ and a typical power of $25\,\text{mW}$ is applied.
    All mentioned beam waists denote $1/e^2$ radii.
    The velocity of the atoms is projected in the direction perpendicular to the optical molasses. 
    The deflection angle was optimized through simulation to sufficiently separate fast ($\gtrsim 66\,\text{m/s}$) and slow ($\lesssim 55\,\text{m/s}$) atoms while ensuring that about $85\,\text{\%}$ of slowed atoms reach the science chamber.
    Because the optical molasses is at a $20\,\degree$ angle with the incoming atomic beam, its velocity is reduced by an additional $(1 - \cos20\,\degree)\sim 6\ \%$ by projecting its velocity onto the new propagation direction.

    The high-velocity atoms are dumped on a heated sapphire window that transmits the Zeeman laser. 
    The window is heated to $200\,\text{°C}$ to maintain sufficient Sr vapor pressure for re-evaporation onto the colder chamber walls and to avoid the window becoming opaque from Sr deposition.
    Glass wool insulation and a surrounding water-cooled housing limit heat transfer to the science chamber.

\subsection{Science Chamber and Magnetic Field Control}\label{subsec:science_chamber}
    
\begin{figure}
    \centering
    \includegraphics[width=\linewidth]{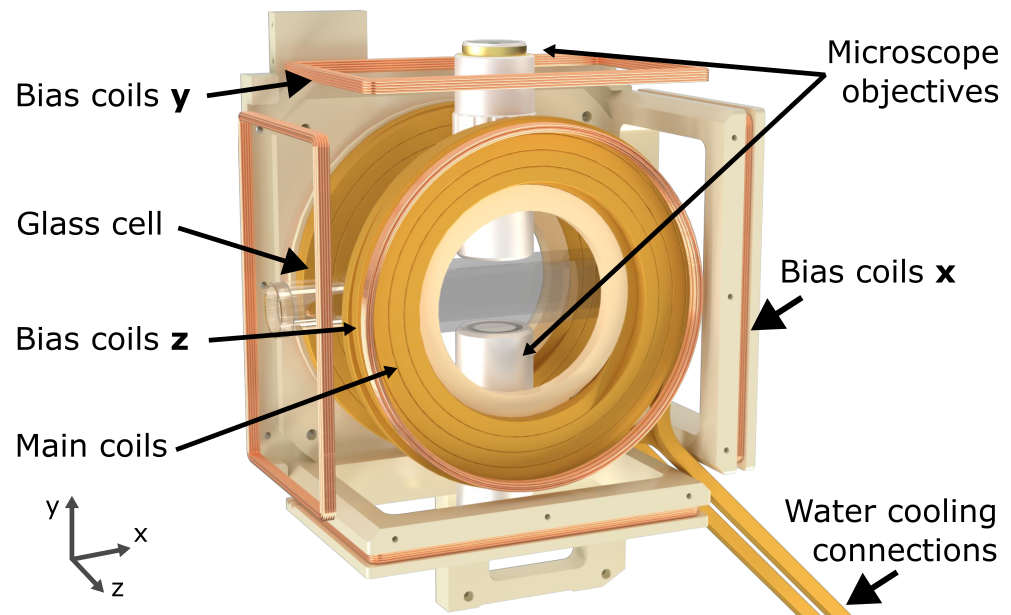}
    \caption{
        The science chamber (glass cell) with surrounding magnetic field coils and microscope objectives. 
        The Sr atoms enter the cell along the $\textbf{x}$ direction.
        The central thick copper winding represents the main coil, constructed from hollow, water-cooled wire, and employed for generating MOT and Helmholtz magnetic fields.
        Three pairs of bias coils corresponding to the Cartesian coordinates \textbf{x}, \textbf{y} and \textbf{z}, depicted in orange, produce offset fields. 
        These coils are wound and mounted onto the coil holder (beige).
        Detailed parameters for all coils are summarized in \cref{tab:coils}.
        The layout of the \textbf{x} bias coils allow the diagonal MOT beams ($5\,\text{mm}$ $1/e^2$ radius waist) to enter the glass cell at a $25\,\text{\degree}$ angle w.r.t the \textbf{x}-axis (see \cref{fig:optics_setup}) to make room for the microscope objectives. 
    }
    \label{fig:coils}
\end{figure}

    After a pumping section, the slowed atoms enter the science chamber, where the atoms are cooled and trapped to create an array of single atoms.
    The core of the science chamber, as illustrated in \cref{fig:coils}, consists of an optically contacted fused silica glass cell\equip{Japan Cell Co., Ltd.} surrounded by two microscope objectives \equip{Mitutoyo, G Plan APO $50\times$}, a pair of main coils and three pairs of compensation coils. 
    The geometry of the science chamber was designed with optical access in mind.
    Four of the six MOT beams (diagonal beams) enter the glass cell under a $25\,\text{\degree}$ angle with respect to the horizontal to make room for the microscope objectives.
    The other two MOT beams (horizontal beams) propagate along the $\mathbf{z}$-axis.
    
    Good vacuum conditions are obtained with a second differential pumping stage (length $150\,\text{mm}$, inner diameter $22.1\,\text{mm}$), leading to a conductance of $19\,\text{L/s}$ and an ion pump and non-evaporative getter combination \equip{Saes Getters, D500 StarCell}.
    Given the design of the vacuum chamber and the pumps used, we expected to reach a pressure of low $10^{-11}\,\text{mbar}$ in the science chamber.
    This matches the pressure that we read from the ion pumps. 
    A measurement of the MOT lifetime yielded a pressure of $3\times10^{-11}\,\text{mbar}$ when comparing the lifetime with that of other groups \cite{snigirev_fast_2019}.

    The magnetic field coils surrounding the glass cell are mounted on a custom-designed holder fabricated from polyether ether ketone (PEEK), specifically chosen to prevent eddy currents, which could reduce the response speed of the current-driving system.
    The coil parameters are summarized in \cref{tab:coils}.

\begin{table*}[t]
    \centering
    \caption{
        Measured parameters of the main and bias coils.
        The dimensions of the main coils refer to the inner diameter of the innermost winding.
        The wire gauge is specified as the cross-sectional dimensions with outer size (hollow inner size) indicated.
        Resistance and inductance are measured for each pancake of the coil.
        The time constant is calculated as inductance over resistance.
    }
    \begin{ruledtabular}
    \resizebox{0.9\textwidth}{!}{%
    \begin{tabular}{lcccc}
                                            & \textbf{Main coil}       & \textbf{Bias coil y}  & \textbf{Bias coil z}  & \textbf{Bias coil x} \\ \hline
        Coil dimensions (mm)                & 92                       & 87$\times$140         & 140                   & 87$\times$140 \\
        Coil separation (mm)                & 47                       & 172                   & 88                    & 172 \\
        Wire gauge (mm)                     & 5$\times$5 (hollow 3)    & 1                     & 1                    & 1 \\
        Windings per coil                   & 8                        & 20                    & 20                    & 20 \\
        Resistance ($\Omega$)               & 0.003                    & 0.415                 & 0.425                 & 0.420 \\
        Inductance ($\mu$H)                 & 5.4                      & 111.3                 & 122.9                 & 118.9 \\
        Time constant (ms)                  & 1.83                     & 0.27                  & 0.29                  & 0.28 \\
        Magnetic field strength, Helmholtz (G/A) & 1.23                     & 0.548                 & 1.625                 & 0.696 \\
        Gradient field strength, anti-Helmholtz (G/cm/A)             & 0.235                    & n.a.                  & n.a.                  & n.a. \\
    \end{tabular}
    }
    \end{ruledtabular}
    \label{tab:coils}
\end{table*}

    The main coils are used to generate a gradient field for MOT operation (anti-Helmholtz configuration) and a Helmholtz field to drive the clock transition \cite{taichenachev_magnetic_2006} and consist of two pancake-shaped structures, each comprising two layers with four windings per layer.
    All windings follow the so-called $\alpha$-type configuration, minimizing unwanted fields generated by the leads. 
    Additionally, the two pancake coils are arranged as mirror images to further reduce field inhomogeneities when operating in Helmholtz mode.
    The coils are hollow for water cooling purposes, dissipating about 240 W for maximum current.
    A chiller maintains the water temperature within $\pm1\,\text{\degree C}$ for magnetic field stability.

    The main coil is powered by a primary power supply \equip{Delta Electronika, SM30-200} that provides up to $200\,\text{A}$.
    It features a noise and ripple of about $100\,\text{ppm}$, corresponding to $20\,\text{mG}$ magnetic field fluctuations at $200\,\text{G}$.
    These fluctuations limit the expected single-qubit gate fidelity, but could be mitigated by current-feedback stabilization \cite{borkowski_active_2023}.
    To produce both Helmholtz and anti-Helmholtz fields with a single set of coils, we use an H-bridge of four water-cooled metal–oxide–semiconductor field-effect transistors to control the current direction in one of the coils.
    The low internal resistance of the primary power supply prevents rapid quenching of the current needed to switch from the blue to the red MOT, operating on the narrow $689\,\text{nm}$ intercombination line \cite{katori_magneto-optical_1999}.
    Therefore, a secondary power supply \equip{Delta Elektronika, SM70-AR-24} was added in parallel, and a second H-bridge circuit is used for quickly switching off the primary power supply, leaving only the secondary (low-current) power supply output.
    Using a flux gate sensor \equip{LEM, IN 400-S}, the measured switching speed of the current (gradient magnetic field strength) from $195\,\text{A}$ ($46\,\text{G/cm})$ to $5.5\,\text{A}$ ($1.3\,\text{G/cm}$) is $\sim0.1\,\text{ms}$.
    
    The bias coils, used for the compensation of stray magnetic fields and the movement of the MOT quadrupole field center in 3-D space, consist of three pairs of coils aligned along the $x$, $y$, and $z$ axes (see \cref{fig:coils} and \cref{tab:coils}).
    The y-axis bias coils are mounted onto the main coils and satisfy the Helmholtz condition. 
    The x- and z-axis bias coils are attached to the sides of the main coil holder, with a separation greater than that required for the Helmholtz condition.
    However, their geometry still provides good magnetic field uniformity for small applied fields of maximum $\sim 0.5\,\text{G}$ over the size of a typical tweezer array of $\mathcal{O}(100)\,\text{$\mu$m}$.
    The coils are driven by power supply units \equip{Delta Elektronika, ES015-10}, selected for their low noise and ripple specifications. 
    This ensures that the resulting magnetic field noise and drift remain below the mG level.

\subsection{Control System}

    The timing sequence for the magnetic field control and the laser beams used in the experiment is provided by our control system.
    The core of the control system is the open-source software platform ARTIQ \cite{bourdeauducq_artiq_2016} and the open-source hardware SINARA \cite{noauthor_sinara_2023}.
    Designed to be used on top of ARTIQ, a flexible abstraction layer has been written, which can be found in Ref. \cite{delpozomellado2025}.

\section{Producing Single Sr Atoms}\label{sec:single_atoms} 

    To realize an array of single Sr atoms, we cool and trap them using several MOT stages within the science chamber (\cref{subsec:mots}).
    Subsequently, we overlap them with the tweezer array and prepare and image single atoms (\cref{subsec:tweezer_imaging}).
    Finally, the tweezers are characterized (\cref{subsec:tweezer_charac}).

\subsection{MOT Operation and Characterization}\label{subsec:mots}

    The atoms are cooled and trapped using a blue MOT followed by a red MOT.
    The blue MOT beams are generated by splitting the $461\,\text{nm}$ laser output into six fiber outputs using a fiber splitter\equip{Evanescent Optics Inc., Custom 1x6 PM splice-less coupler array with input tap}.
    The $689\,\text{nm}$ laser light is split with a fiber port cluster\equip{Schäfter+Kirchoff GmbH, 47-FPC-1-6-689}, allowing flexible adjustment of the individual output powers.
    To minimize the amount of optical access required, the $461\,\text{nm}$ and $689\,\text{nm}$ MOT beams are combined together using dichroic mirrors. 
    In addition, three repump laser beams at $679\,\text{nm}$ (\termSym{3}{P}{0} $\to$ \termSym{3}{S}{1}), $688\,\text{nm}$ (\termSym{3}{P}{1} $\to$ \termSym{3}{S}{1}) and $707\,\text{nm}$ (\termSym{3}{P}{2} $\to$ \termSym{3}{S}{1}) are combined with the horizontal red MOT beams using a fiber multiplexer (Evanescent Optics Inc., 4x4 splice-less PM coupler array).

\begin{figure*}
    \centering
    \includegraphics[width=0.94\textwidth]{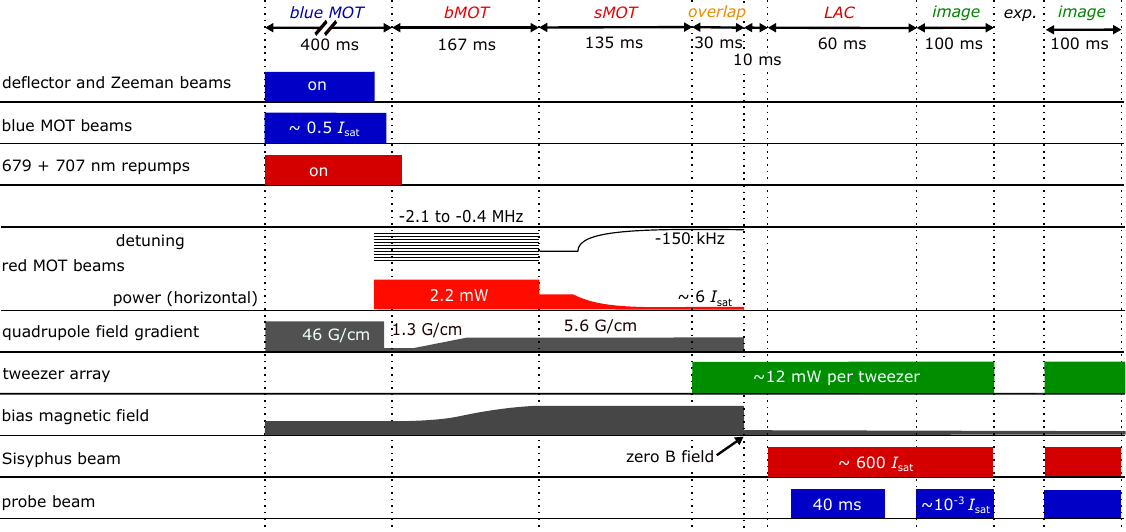}
    \caption{
        Overview of the experimental time sequence used to load and image the tweezer array. 
        The atoms are cooled using the blue MOT, the broadband red MOT (bMOT) and the single-frequency red MOT (sMOT). 
        The sMOT is overlapped in space with the tweezer array using the bias magnetic field.
        Prior to imaging, light-assisted collisions (LAC) and cooling in tweezers are performed using the Sisyphus beam, accelerated using the blue probe beam. 
        Two subsequent fluorescence images are taken with an exposure time of $100\,\text{ms}$. 
        In between the two images, tweezer characterization experiments (\textit{exp.}) are performed. 
    }
    \label{fig:exp_sequence}
\end{figure*}

    First, atoms are captured and cooled using the blue MOT operating on the broad \termSym{1}{S}{0} $\to$ \termSym{1}{P}{1} transition, which has a linewidth of $30.5\,\text{MHz}$ \cite{yasuda_photoassociation_2006}.
    The detuning of the beams is roughly one linewidth, the gradient field strength is $46\,\text{G/cm}$ and the intensity is $\sim 0.5\,I_{\text{sat}}$.
    During blue MOT loading, the $679\,\text{nm}$ and $707\,\text{nm}$ repump laser are shone on the atoms to prevent them from being lost in the cooling cycle due to the \termSym{1}{P}{1} $\to$ \termSym{3}{P}{2} decay channel \cite{yasuda_lifetime_2004}.
    The time sequence used is sketched in \cref{fig:exp_sequence}.
    The blue MOT loading time used $t_{\text{load}}$ is typically $400\,\text{ms}$, as this is enough to uniformly load the tweezer array.

\begin{figure}
  \centering
  \includegraphics[width=\linewidth]{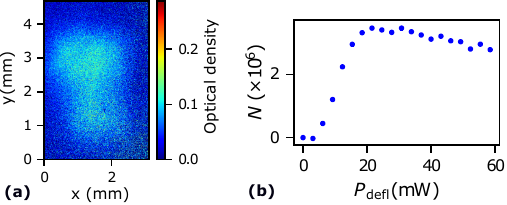}
  \caption{
    Blue MOT operation.
    a) In-situ absorption image for a deflector beam power $P_{\text{defl}}=25\,\text{mW}$.
    b) Number of atoms in the blue MOT $N$ as a function of $P_{\text{defl}}$. 
    The loading time used is $t_{\text{load}}=1500\,\text{ms}$.
  }
  \label{fig:nr_atoms_blue}
\end{figure}

    In \cref{fig:nr_atoms_blue}a, an in-situ absorption image is shown for $t_{\text{load}}=1500\,\text{ms}$ and a deflector beam power of $P_{\text{defl}}=25\,\text{mW}$.
    The shape of the blue MOT depends on the alignment of the laser beams because of the power dependent scattering force as the transition is not saturated, but this has no effect on subsequent stages of the experiment.
    From the optical density of the absorption image, the atom number $N$ is estimated, which is plotted in \cref{fig:nr_atoms_blue}b and strongly depends on the deflector laser power $P_{\text{defl}}$, which is typically operated at $25\,\text{mW}$.
    For lower powers, $N$ decreases and for powers higher than $P_{\text{defl}}\gtrsim 30\,\text{mW}$ or $I\gtrsim2\,I_{\text{sat}}$ also a lower atom number is observed.
    The latter is most likely because more atoms are lost from the cooling cycle, as no repump light is present in the deflector laser beams.

    Second, the atoms are further cooled on the $7.4\,\text{kHz}$ \termSym{1}{S}{0} $\to$ \termSym{3}{P}{1} red line following the time sequence sketched in \cref{fig:exp_sequence}.
    We achieve a $\sim30\,\text{\%}$ transfer efficiency from the blue to the red MOT following the method of Ref. \cite{katori_magneto-optical_1999}.
    Before the blue beams are switched off, the red beams are on for $40\,\text{ms}$ at full power ($2.2\,\text{mW}$ measured in the horizontal beams), while the atomic beam is switched off by switching off the deflector and ZS laser beams. 
    The red MOT beams have a waist of $3.2\,\text{mm}$.
    The diagonal beams have roughly twice as much power as the horizontal beams, to compensate for the extra gravity force in this direction. 

    To increase the transfer efficiency, the red MOT is applied first in broadband mode (bMOT) and then in single-frequency mode (sMOT). 
    In bMOT mode, the red MOT beams are modulated using $1.7\,\text{MHz}$ frequency modulation depth and a  $45\,\text{kHz}$ modulation frequency, and the quadrupole field gradient changed rapidly to $1.3\,\text{G/cm}$ (\cref{subsec:science_chamber}).
    Bias fields are applied to spatially overlap the two MOTs.
    Subsequently, the gradient field is gradually increased again to compress the bMOT.
    This increases the density and reduces the sensitivity to stray magnetic fields when overlapping with the tweezer array.
    By applying smooth, minimal-jerk trajectories to the magnetic field using the bias coils, the bMOT position is shifted to the focal point of the microscope objective where the tweezers array is located.
    After the broadband phase, cooling continues in the sMOT stage by disabling frequency modulation.
    The intensity (horizontal beams) and the detuning are exponentially reduced (\cref{fig:exp_sequence}) to their final values of $\sim 6\,I_{\text{sat}}$ and $-150\,\text{kHz}$.
    These final settings are held constant for $30\,\text{ms}$ to give the MOT time to reach thermal equilibrium.

\begin{figure}
    \centering
    \includegraphics[width=\linewidth]{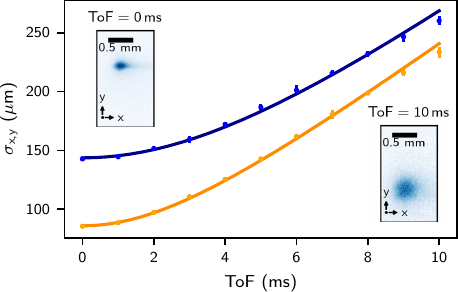}
    \caption{
        Temperature measurement of the sMOT.
        The standard deviations $\sigma_x$ (blue points) and $\sigma_y$ (orange points) obtained from 2-D Gaussian fits of the sMOT cloud are plotted for variable time of flight (ToF) duration.
        Two fluorescence images for $\text{ToF}=0\,\text{ms}$ and $\text{ToF}=10\,\text{ms}$ are shown as well.
        The data points are fit in $x$ (blue solid line) and $y$ (orange solid line) to \cref{eq:time_of_flight} to extract the temperatures $T_x$ and $T_y$.
        Error bars represent the standard error of the mean.
    }
    \label{fig:red_mot_temperature}
\end{figure}

    After the MOT stages, we measure a final temperature of $T_x=T_y = 5(1)\,\text{$\mu$K}$ for the horizontal $x$ and the vertical $y$ directions using time of flight (ToF) fluorescence imaging on the blue transition, as shown in \cref{fig:red_mot_temperature}.
    After release, images are acquired by applying resonant $461\,\text{nm}$ pulses of $50\,\text{$\mu$s}$ duration.
    The images are fit to 2-D Gaussian functions with standard deviations in the $x$ and $y$ directions $\sigma_{x,y}$, which are subsequently fit to the function \cite{arora_measurement_2012}

    \begin{equation}\label{eq:time_of_flight}
        \sigma_i^2(\text{ToF})=\sigma_{i,0}^2 + \frac{k_B T_i}{m} \text{ToF}^2,
        \quad i=\{x,y\}
    \end{equation}
    with $\sigma_{x_0},\sigma_{y_0},T_x$ and $T_y$ as fit parameters.
    The reported error is a combination of statistical and systematic errors, but is dominated by a systematic error in the magnification of the imaging system used.  
    Images for $\text{ToF}=0\,\text{ms}$ and $\text{ToF}=10\,\text{ms}$ show a simultaneous drop due to gravity and an expansion due to finite temperature \cite{katori_magneto-optical_1999}.
    The number of atoms $N$ in the sMOT for $1500\,\text{ms}$ of blue MOT loading time is determined using an absorption measurement that yields $N \sim 3 \times 10^5$.

\subsection{Tweezer Production and Imaging}\label{subsec:tweezer_imaging}
\begin{figure}
        \centering
        \includegraphics[width=\linewidth]{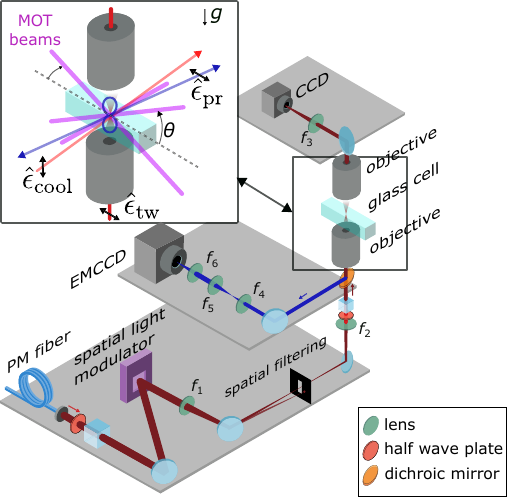}
        \caption{
            Schematic optical setup used for the tweezer production and imaging of single atoms in tweezers. 
            The tweezer laser (dark red) reflects off a spatial light modulator and is projected into the glass cell using a telescope with magnification $f_2/f_1$ and the bottom objective.
            The tweezer pattern is imaged onto a CCD camera using the top objective and the focusing lens $f_3$. 
            Atomic fluorescence (blue laser path) is collected using the bottom objective, separated using a dichroic mirror and sent onto the EMCCD camera using a relay telescope with magnification $f_5/f_4$ and field lens $f_6$.
            Inset: laser beam geometry and quantization axis.
            The MOT beams are indicated in purple and contain $461\,\text{nm}$ and $689\,\text{nm}$ light.
            The diagonal MOT beams have an angle $\theta=25\,\text{\degree}$ with the horizontal axis. 
            The tweezer polarization $\hat{\epsilon}_{\text{tw}}$ sets the quantization axis (dashed line). 
            The probe beam (blue) is retro-reflected and horizontally polarized ($\hat{\epsilon}_{\text{pr}}$). 
            A cross-section of the 3-D dipole emission pattern is indicated using the blue figure 8 curve. 
            The non-retroreflected Sisyphus cooling laser is vertically polarized ($\hat{\epsilon}_{\text{cool}}$). 
        }
        \label{fig:optics_setup}
    \end{figure}

    The tweezers are produced using a spatial light modulator (SLM) and a high-resolution microscope objective, as depicted in detail in \cref{fig:optics_setup}.
    The $813\,\text{nm}$ laser power is controlled in intensity and stabilized using a double-pass AOM and subsequently sent to the experiment table using a high-power polarization-maintaining optical fiber\equip{OZ Optics.}
    The light is collimated to and reflected off of an SLM\equip{Meadowlark, E-series $1920\times1200$}. 
    The $1/e^2$ radius of $4.8\,\text{mm}$ matches the short semi-width of the rectangular SLM.
    The phase mask on the SLM consists of 4 contributions \cite{nogrette_single-atom_2014}.
    These include a wavefront-correction phase, to correct for imperfections in the flatness of the SLM as well as a mask that generates a $5\times5$ spot array in the focus of the objectives using the weighted Gerschberg-Saxton algorithm \cite{gerchberg_practical_1972}.
    Furthermore, a grating phase is applied to separate the zeroth- and first diffraction orders.
    Finally, a Zernike phase mask, consisting of 11 Zernike polynomials, is applied to counteract optical aberrations.
    Polynomials are re-orthonormalized for the rectangular geometry of the SLM \cite{mahajan_orthonormal_2007} and centered around the middle of the SLM pixel array.
    The method of finding the Zernike phase mask is explained in \cref{sec:appendix}.

    The SLM plane is conjugated \cite{nogrette_single-atom_2014} with the pupil of the microscope objective using a Keplerian telescope with a magnification given by the ratio of focal lengths $f$ of $f_2/f_1=0.4$.
    The zero-order light is filtered out in the intermediate focus plane. 
    The objective can be aligned perpendicularly to the vacuum cell using a 5-axis motorized stage\equip{Newport, 8081-M}.
    The total tweezer power is typically $\sim300\,\text{mW}$, or $12(1)\,\text{mW}$ per trap when a $5\times5$ spot array is used.
    The tweezer array projected into the glass cell is imaged by an identical objective onto a CCD camera\equip{FLIR, BFS-31S4M-C}.

    The timing sequence for loading the tweezer array is shown in \cref{fig:exp_sequence}.
    After the sMOT has been moved and given $30\,\text{ms}$ of time to equilibrate, the laser creating the tweezer array is turned on and after $30\,\text{ms}$ of overlap time, the sMOT beams and the gradient field are switched off.
    At the moment the sMOT gradient field is switched off, the magnetic field at the position of the atoms is set to zero using the bias coils, making the linear tweezer polarization axis the proper choice for a quantization axis \cite{urech_narrow-line_2022}.

    In order to prepare a single atom per site, an attractive Sisyphus cooling process is used, which induces light-assisted collisions (LAC) and cools the atoms in the tweezers \cite{schlosser_sub-poissonian_2001, cooper_alkaline-earth_2018,covey_2000-times_2019}. 
    This method requires the excited state to have a larger polarizability than that of the ground state.
    For $813\,\text{nm}$ tweezer light, this condition is met for \termSym{3}{P}{1}$(m_\text{j}=\pm1)$ which has $1.24\times$ higher polarizability than \termSym{1}{S}{0} \cite{urech_narrow-line_2022}.
    The cooling transition is addressed using a single non-retro-reflected vertically polarized Sisyphus beam ($\hat{\epsilon}_{\text{cool}}$ in inset of \cref{fig:optics_setup}).
    During the LAC stage, the detuning of the Sisyphus cooling beam is set to $-2.64\,\text{MHz}$ to compensate for the differential AC Stark shift of the tweezers and the intensity is $\sim 600\,I_{\text{sat}}$ at the location of the atoms.
    The LAC process is accelerated by adding the blue probe laser \cite{wen_apparatus_2024} for $40\,\text{ms}$.
    The probe laser, which is also used for MOT imaging, has a detuning of $-100\,\text{MHz}$ with respect to the free-space resonance and an intensity of $\sim 10^{-3}\,I_{\text{sat}}$.
    In the inset of \cref{fig:optics_setup} the horizontal polarization of the probe laser is indicated with the vector $\hat{\epsilon}_{\text{pr}}$.
    Its polarization aligns with the polarization of the tweezers $\hat{\epsilon}_{\text{tw}}$ ($\pi$-transition).

    During imaging, the Sisyphus and probe beams are used for $100\,\text{ms}$ and the collected atomic fluorescence at $461\,\text{nm}$ is separated from the tweezers beam using a long-pass dichroic filter\equip{Thorlabs, $505\,\text{nm}$ cut-on}, and relayed onto an electron multiplying charge coupled device (EMCCD) camera \equip{Andor, iXon 888 Ultra} using a telescope of magnification $f_5/f_4=1.2$ and field lens $f_6=50\,\text{mm}$.

    We optimize survival probability during imaging by scanning the Siyphus cooling laser as shown in \cref{fig:histogram}c.
    The survival probability is found by taking two consecutive images. 
    The highest survival probability was found to be $0.99^{+0.01}_{-0.02}$ for a detuning of $-2.7\,\text{MHz}$, with respect to the free-space resonance. 
    The estimated AC Stark shifted resonance of $-2.35\,\text{MHz}$ is indicated in the vertical red line and is further detailed in the following \cref{subsec:tweezer_charac}.

    The average atomic fluorescence image we obtain for a $5\times5$ spot array is shown in \cref{fig:histogram}a and the histogram of the computed photon counts for individual regions of interest (RoI) is shown in \cref{fig:histogram}b.
    A clear bimodal distribution can be seen, indicating single atom preparation.
    The histogram was obtained from 25 RoIs in 484 experimental runs.
    Each RoI has its own weighting matrix of $5\times5$ pixels, where the weights are obtained from the average image \cite{manetsch_tweezer_2025}.
    The binary threshold to distinguish the presence of an atom was chosen to maximize the probability of correctly labeling the presence of an atom \cite{madjarov_entangling_2021, nelson_imaging_2007}, which yields $5.67$ photons as a threshold.
    Using this threshold, we observe a filling fraction (probability of occupying the trap) of $\sim46$ \% and imaging fidelity (probability of correctly identifying the presence of an atom) of $\sim 0.997$.

\subsection{Tweezer Characterization}\label{subsec:tweezer_charac} 

\begin{figure}
    \centering
    \includegraphics[width=\linewidth]{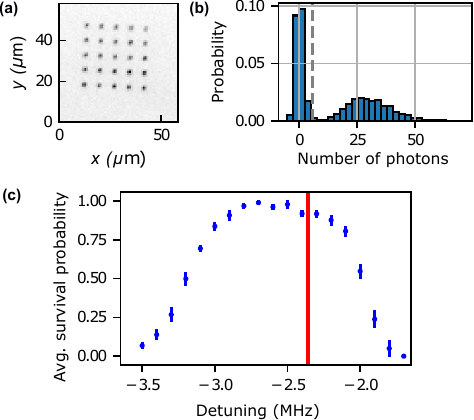}
    \caption{
        Single-atom imaging and thresholding.
        a) Average image on the EMCCD of 484 experimental runs.
        b) Histogram of the number of measured photons per region of interest (RoI) of 484 images and the chosen detection threshold of $5.67$ photons (dashed gray line).
        c) Array-averaged survival probability during imaging as a function of the detuning of the Sisyphus cooling laser (blue points) and estimated Stark-shifted resonance at $-2.35\,\text{MHz}$ (red line).
        Error bars represent the standard error of the mean. 
    }
    \label{fig:histogram}
\end{figure}

\begin{figure}[t]
    \centering
    \includegraphics[width=\linewidth]{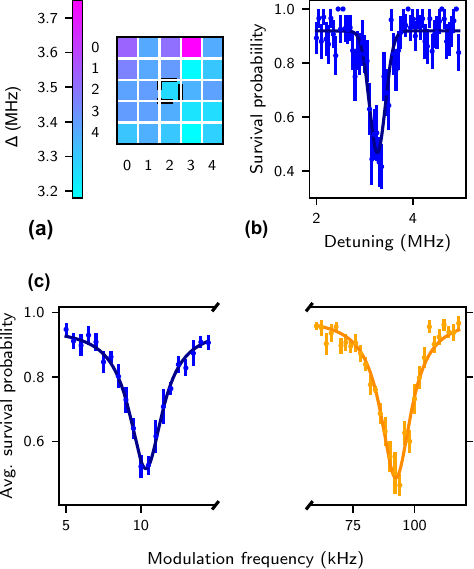}
    \caption{
        Characterization of the tweezers.
        a) Uniformity of the differential AC Stark shift $\Delta$ of the array, showing the location of the Gaussian loss feature per array site for the array indices numbered 0-4 in the horizontal and vertical directions.
        b) Survival probability (blue dots) for a single array site, the middle site indicated with the black dashed lines in (a), after applying a blue-detuned Sisyphus cooling laser for $50\,\text{ms}$ in between two images as a function of Sisyphus laser detuning.
        The Gaussian fit (blue solid line) is shown, which was used to calculate $\Delta$ in a).
        c) Results of the parametric heating measurement to calculate the average trap frequencies of $5.74(2)\,\text{kHz}$ (longitudinal) and $51.3(2)\,\text{kHz}$ (radial).
        The average survival probability over the array as a function of modulation frequency is shown. 
        The datasets corresponding to the longitudinal trapping direction (blue points) and the axial direction (orange points) are plotted and fit with Lorentzian functions (blue and orange solid lines) centered at $10.33(4)\,\text{kHz}$ and $92.3(3)\,\text{kHz}$ respectively. 
        Error bars represent the standard error of the mean.
    }
    \label{fig:tweezer_characterization}
\end{figure}

    Using single atom imaging, we characterize the depth of individual tweezers using loss spectroscopy, as shown in \cref{fig:tweezer_characterization}a and \cref{fig:tweezer_characterization}b.
    We follow the experimental sequence shown in \cref{fig:exp_sequence}, inserting a block $50\,\text{ms}$ between the images, where only the Sisyphus laser is shone on the atoms with variable detuning.
    Because the \termSym{3}{P}{1}$(m_\text{j}=0)$ state has a lower polarizability than the ground state, positive detuning of the Sisyphus beam can induce heating, leading to a drop in survival probability. 
    We determine the resonance center of this loss feature by fitting a Gaussian to the survival data \cite{urech_narrow-line_2022}.
    This center, measured at an average of $\Delta=+3.37(2)\,\text{MHz}$, we assume corresponds to the differential AC Stark shift between the ground and \termSym{3}{P}{1}$(m_\text{j}=0)$ states.
    From the ratio of polarizabilities of these two states, this translates to an AC Stark shift for the ground state $U_0$ of $U_0=h\times11.08(7)\,\text{MHz}$ or $k_B\times 0.532(3)\,\text{mK}$ for the data in \cref{fig:tweezer_characterization}a and \cref{fig:tweezer_characterization}b.

    Note that the trap depth $U_0$ also influences our Sisyphus cooling procedure. 
    We can use this result to estimate the location of the AC Stark shifted resonance for the \termSym{1}{S}{0} $\to$ $\termSym{3}{P}{1}(m_{\text{j}}=\pm1)$ cooling transition from the ratio of polarizabilities, yielding $-2.35\,\text{MHz}$ which is indicated by the red line in \cref{fig:histogram}c.
    We see that the survival probability is indeed the highest when slightly red-detuned from the AC Stark shifted resonance, corresponding to an attractive Sisyphus cooling process \cite{urech_narrow-line_2022, covey_2000-times_2019}.

    \Cref{fig:tweezer_characterization}a shows the variation in $\Delta$ as measured across the $5\times5$ array. 
    We obtain this uniformity after several optimization steps following the method of Ref. \cite{nogrette_single-atom_2014} where we adjust the phase mask of the SLM by adjusting the target weights for the Gerchberg-Saxton algorithm.
    After 3 rounds of updating the phase mask, the calculated standard deviation in the trap depth reaches $3.4\,\text{\%}$ and no further improvement was observed.

    In \cref{fig:tweezer_characterization}c, results from the parametric heating method \cite{savard_laser-noise-induced_1997} used to compute trap frequencies are shown for the longitudinal (blue) and radial (orange) trapping directions.
    The measurement was performed by first adiabatically lowering the trapping laser power \cite{de_leseleuc_quantum_2018} to a factor $\chi=0.21$ (longitudinal) or $\chi=0.35$ (radial) of the maximum trapping power corresponding to a trap depth $U_0$.
    Following this reduction, we introduce a sinusoidal modulation of the trapping power and vary the modulation frequency.
    At a modulation frequency of twice the trap frequency, the survival probability is expected to drop, assuming harmonic traps \cite{savard_laser-noise-induced_1997}.
    In shallower tweezers, the observed loss feature is more pronounced. 
    Finally, the traps are ramped up to $U_0$ again and a survival image is taken. 
    The drops in survival probability in \cref{fig:tweezer_characterization}c are assumed to correspond to the longitudinal and radial directions, respectively, and are centered around $10.33(4)\,\text{kHz}$ and $92.3(3)\,\text{kHz}$.
    From the location of the loss features, the trap frequencies can be extracted.
    For Gaussian traps, the atoms explore anharmonic regions of the potential, resulting in a resonance condition shifted to a factor of $\sim 1.8$ instead of 2 (harmonic potential) \cite{jauregui_nonperturbative_2001}.
    Taking this into account, the trap frequencies in shallower traps of depth $\chi U_0$ are $5.74(2)\,\text{kHz}$ (longitudinal) $51.3(2)\,\text{kHz}$ (radial).
    The trap frequencies at maximum trap depth $U_0$ are by multiplying with the factor $\chi^{-1/2}$, which yields $12.7(4)\,\text{kHz}$ (longitudinal) and $86.4(3)\,\text{kHz}$ (radial).
    The quoted uncertainties are derived from the fits only and do not include systematic uncertainties associated with the anharmonic correction factor or the power scaling.

    Finally, from the combination of the trap depth and trap frequency measurements, the average tweezers waist $w_0$ can be calculated yielding $w_0=0.81(2)\,\text{$\mu$m}$, which is slightly above the diffraction-limit of $0.75\,\text{$\mu$m}$ set by the microscope objective \cite{chon_dependence_2007}.
    We anticipate that the measured trap depth, trap frequencies, waist, and uniformity satisfy the performance criteria necessary to implement qubit manipulation.

\section{Conclusions and Outlook}

    In conclusion, we have detailed the design and construction of a versatile Sr tweezer apparatus featuring a custom designed oven, Zeeman slower and a deflection stage.
    We characterized the multi-stage MOT operation and demonstrated the steps involved to achieve single atoms in tweezers. 
    Performance is validated by consistent preparation of single Sr atoms in a $5\times5$ array of optical tweezers that feature a survival probability of $0.99^{+0.01}_{-0.02}$ and an imaging fidelity of $\sim0.997$.
    The tweezers were optimized in terms of trap depth and uniformity.
    Characterization measurements yield a waist of $w_0=0.81(2)\,\text{$\mu$m}$, which is slightly above the diffraction limit. 

    With this, the atomic array platform presented here can form the basis for an atomic quantum processor with the aim of performing near-term variational algorithms \cite{de_keijzer_pulse_2023,peruzzo_variational_2014}.
    To this end, we are implementing a movable optical tweezer for atomic re-arrangement \cite{barredo_atom-by-atom_2016}, the $698\,\text{nm}$ laser system for coherent manipulation of the $\termSym{1}{S}{0}$ $\leftrightarrow$ $\termSym{3}{P}{0}$ clock transition \cite{norcia_microscopic_2018}, and the $317\,\text{nm}$ laser for targeting the $(5s61s) \termSym{3}{S}{1}$ Rydberg state \cite{madjarov_high-fidelity_2020}.
    We are also working towards full-stack integration with the openly accessible Quantum Inspire platform \cite{noauthor_qutech_2025,noauthor_ry_2025} and the Rydberg emulator RySP \cite{noauthor_tues_2025}.
    To scale up the number of tweezer sites, we plan to replace the microscope objective with an objective designed to have better transmission at \qtyunit{813}{nm}.
    Using the full \qtyunit{813}{nm} laser power and \qtyunit{0.5}{mK} trap depths, we would reach about 70 sites. 
    While laser power ultimately limits the number of sites, we anticipate that adding a shallow-angle lattice trap \cite{hofer_single-atom_2025, young2022} to boost the longitudinal trapping frequency would allow us to reduce the tweezer trap depth while maintaining high imaging survival probability, relaxing the per-site power requirement.
    
    


\begin{acknowledgments}
    
The authors thank the entire neutral-atom KAT-1 collaboration  (\href{https://www.tue.nl/en/research/research-groups/center-for-quantum-materials-and-technology-eindhoven/projects/rydberg-atom-quantum-computing-and-simulation}{www.tue.nl/rydbergQC}), especially the RuBy and theory teams at TU/e.
We thank our technicians E. Rietman, H. van Doorn, H. van den Heuvel and S. Oosterink for technical assistance and the Eindhoven Prototyping Center and especially P. de Laat for aid with the design and manufacturing of the apparatus.
Finally, we thank master students R. Teunissen and G. Kucharska for contributions to the experiment.
This work received funding from the Dutch ministry of economic affairs and climate policy (EZK), as part of the Quantum Delta NL program, from the Netherlands Organization for scientific Research (NWO) under grant no. 680.92.18.05 and no. QDNL/NWO AL-1 project number NGF.1623.23.025 and from the the Horizon Europe program HORIZON-CL4-2021-DIGITAL-EMERGING-01-30 via the project 101070144 (EuRyQa).

\end{acknowledgments}

\section*{Author Declarations}

\subsection*{Conflict of Interest}

    The authors have no conflict of interest to disclose.

\subsection*{Author Contributions}
    
    \textbf{Marijn Venderbosch}: Writing - original draft (lead); Methodology (equal); Investigation (equal); Software (supporting).
    \textbf{Rik van Herk} and \textbf{Zhichao Guo}: Methodology (equal); Software (equal); Investigation (equal); Writing - original draft (supporting).
    \textbf{Jésus del Pozo Mellado}: Software (equal); Methodology (supporting); Investigation (supporting).
    \textbf{Max Festenstein}: Methodology (supporting); Software (supporting); Investigation (supporting); Writing - original draft (supporting);
    \textbf{Deon Janse van Rensburg}: Methodology (supporting); Investigation (supporting);
    \textbf{Ivo Knottnerus}, \textbf{Yu Chih Tseng} and \textbf{Alexander Urech} and \textbf{Robert Spreeuw}: Methodology (supporting).
    \textbf{Florian Schreck}: Conceptualization (supporting); Funding acquisition (supporting); Supervision (supporting).
    \textbf{Rianne Lous}: Writing - review \& editing (lead); Supervision (equal), Methodology (supporting).
    \textbf{Edgar Vredenbregt}: Supervision (equal); Writing - review \& editing (supporting); Conceptualization (supporting); Methodology (supporting).
    \textbf{Servaas Kokkelmans}: Conceptualization (lead); Funding acquisition (lead); Supervision (equal).

\section*{Data Availability}

    The data that support the findings of this study are openly available in the 4TU database at Ref. \cite{venderbosch_data_2026}.
    Additional data is available from the corresponding author upon reasonable request.

\bibliographystyle{apsrev4-2}
\bibliography{admin/references}

\appendix
\section{Aberration Correction}\label{sec:appendix}

    The initial aberration correction was performed by minimizing the spot sizes of the tweezers on the CCD camera in \cref{fig:optics_setup}, while individually scanning the Zernike coefficients. 
    This approach assumes that the Zernike mask affects the tweezers globally, which is an approximation.
    Although the Zernike polynomials are mathematically orthogonal, in practice we observed slight coupling between the modes during optimization.
    Possibly, this arises from imperfections in the tweezer projection optical path: for the $813\,\text{nm}$ laser, a Gaussian beam is used instead of uniform illumination, which may be misaligned with respect to the SLM center. 
    Indeed, we have observed that this misalignment is present in our setup. 
    Furthermore, imperfections in the electro-optic calibration that was performed to linearize the SLM phase response may contribute to coupling. 
    
    Subsequently, an in-situ measurement from trapped atoms was used \cite{jenkins_ytterbium_2022,endres_atom-by-atom_2016}.
    While loss spectroscopy on the \termSym{1}{S}{0} $\to$ \termSym{3}{P}{1}$(m_\text{j}=0)$ transition (\cref{fig:tweezer_characterization}b) provides a measurement of the trap depth, it is slow due to the large number of data points required.
    
    To speed up the method and account for coupling between correction terms that may be present, Zernike coefficients 4--15 (Noll index) were optimized simultaneously using the Machine-Learning Online Optimization Package (M-LOOP) \cite{wigley_fast_2016}, employing a differential evolution algorithm.
    The first three coefficients were omitted because they correspond to an irrelevant global phase and a translation in the focus plane.
    The algorithm minimized a cost function defined by the negative of the survival probability.
    As survival probability was optimized and survival probability plateaus near unity, it is no longer a sensitive proxy to trap depth, which was subsequently solved by using a larger red detuning.
    This process of increasing the detuning was repeated twice until the depth of the tweezers was computed to be nearly diffraction-limited. 
    For each iteration of the Zernike phase mask, 10 identical experimental runs were performed, corresponding to $\mathcal{O}(100)$ loaded tweezers for $\mathcal{O}(10)$ loaded tweezers per experimental run.
    Using this method, the nearly diffraction-limited tweezers spots presented in \cref{subsec:tweezer_charac} were obtained.

\end{document}